\documentclass[%
 reprint,
groupedaddress,
 amsmath,amssymb,
 aps,
prl,
longbibliography
]{revtex4-2}

\usepackage{graphicx}
\usepackage{dcolumn}
\usepackage{bm}

\begin{document}

\preprint{APS/123-QED}

\title{Single-Particle mobility edge without disorder}

\author{Donny \surname{Dwiputra}}
 \email{donny.dwiputra@s.itb.ac.id}
 \affiliation{Theoretical Physics Laboratory, Faculty of Mathematics and Natural Sciences, Institut Teknologi Bandung, Jl. Ganesha 10, Bandung 40132, Indonesia}
\author{Freddy P. \surname{Zen}}
\email{fpzen@fi.itb.ac.id}
 \affiliation{Theoretical Physics Laboratory, Faculty of Mathematics and Natural Sciences, Institut Teknologi Bandung, Jl. Ganesha 10, Bandung 40132, Indonesia}
 \affiliation{Indonesian Center for Theoretical and Mathematical Physics (ICTMP), Bandung 40132, Indonesia}

\date{\today}

\begin{abstract}
The existence of localization and mobility edges in one-dimensional lattices is commonly thought to depend on disorder (or quasidisorder). We investigate localization properties of a disorder-free lattice subject to an equally spaced electric field. We analytically show that, even though the model has no quenched disorder, this system manifests an exact mobility edge and the localization regime extends to weak fields, in contrast to gigantic field for the localization of a usual Stark lattice. For strong fields, the Wannier-Stark ladder is recovered and the number of localized eigenstates is inversely proportional to the spacing. Moreover, we study the time dependence of an initially localized excitation and dynamically probe the existence of mobility edge.
\end{abstract}

\maketitle


\textit{Introduction.}---Anderson localization \cite{anderson1958absence}, a universal and extensively studied quantum phenomenon, reveals that the single-particle eigenstates can become exponentially localized in presence of random disorder. In systems beyond two dimensions, an energy-dependent localization transition may occur as a function of disorder strength \cite{abrahams1979scaling,evers2008anderson}, in which mobility edge (ME) appears as an energy threshold $E_\text{ME}$ separating the localized and extended eigenstates. 

In one dimension (1D), when the random disorder is replaced by a quasirandom incommensurate potential (the Aubry-Andr\'e-Harper model \cite{aubry1980analyticity,harper1955single}), the system manifests  the quantum phase transition, albeit with an energy independent transition (no ME), i.e. all localized eigenstates or all extended, depending on disorder strength. In incommensurate lattices, ME can be obtained by slowly varying the on-site potential \cite{sarma1988mobility,sarma1990localization,liu2017fate,liu2017localization}, introducing a long-range hopping \cite{biddle2009localization,biddle2010predicted,deng2019one,saha2019anomalous}, deforming the on-site potential \cite{ganeshan2015nearest}, or by the so-called mosaic lattices \cite{wang2020one,liu2021exact} where the quasiperiodic potential is inlaid with equally spaced sites of zero potential. The existence of ME in low-dimensional open systems enables strong dephasing-assisted quantum transport \cite{dwiputra2021environment}, which can be further enhanced by a periodic driving \cite{dwiputra2020driving,dwiputra2019driving}, current rectification \cite{balachandran2019energy,saha2019particle}, and strong thermoelectric response \cite{yamamoto2017thermoelectricity,chiaracane2020quasiperiodic}.

However, neither random disorder nor quasiperiodic potential is essential for the rise of localized eigenstates. We refer to the lattice in the absence of the random or quasirandomness as disorder-free model. The notion of disorder-free localization can be traced back to the Wannier-Stark lattice \cite{wannier1962dynamics,fukuyama1973tightly,emin1987existence,holthaus1996localization,hartmann2004dynamics}, where a constant electric field applied to the lattice (resulting in a tilted potential) may give rise to exponentially localized states. These eigenstates can be determined exactly as $|m\rangle=\sum_n \mathcal J_{n-m} (2J/F) |n\rangle$ \cite{fukuyama1973tightly}, where $\mathcal J_\nu (z)$'s are Bessel functions of the first kind of hopping strength $J$ and constant force $F$. For $F>2J$, all the eigenstates are localized i.e. there is no energy-dependent ME. It took quarter of a century for experimental evidence to emerge \cite{mendez1988stark,voisin1988observation}, proofing the equally-spaced energy spectrum (Wannier-Stark ladder), and later in recent experiments involving interaction \cite{taylor2020experimental,scherg2021observing,guo2021observation,guo2020stark,morong2021observation,kohlert2021experimental}. Recent works show that in presence of particle interaction, the Stark lattice exhibits robust many-body localization \cite{schulz2019stark,van2019bloch,yao2020many,zhang2021mobility}, which may be destructed by dephasing noise \cite{wu2019bath} or coherent/incoherent drive \cite{bhakuni2020drive,bhakuni2019effect}. However, the existence of ME in noninteracting disorder-free systems is not well-established.

It is natural to ask whether random disorder or quasi-disorder is at all an essential ingredient for a system to manifest ME. In this Letter, we propose a tractable disorder-free 1D Stark lattice in the equally-spaced mosaic model and analytically prove that it exhibits exact ME. Remarkably, the Wannier-Stark ladder is recovered in the localized regime, and the fraction of localized eigenstates is inversely proportional to the lattice spacing parameter. We also study the time-dependent survival probability of an initially localized wave. The experimental realization of this model may follow from the original proposal of mosaic model \cite{wang2020one}, which involves a spin-dependent potential in a chain of ultracold atoms where the odd (even) lattice sites experience nonzero (zero) potential or, alternatively, in a superlattice superimposed by an electric field where one of the layers is neutral.

\textit{Model.}---We consider a 1D tight-binding Hamiltonian with Stark effect in the so-called mosaic lattice \cite{wang2020one} of length $L$ parameterized by an integer $\kappa$,
\begin{equation}\label{model}
	H = -J \sum_{n} (c_n^\dag c_{n+1} + \text{H.c.}) + \sum_{n} \epsilon_n c_n^\dag c_n,
\end{equation}
where
\begin{equation}\label{pot}
 \epsilon_n =
 \begin{cases}
 	Fn\cos\phi, & n=\kappa l, \\
 	0, & \text{otherwise}.
 \end{cases} 
\end{equation}
Here $c_n$ is the annihilation operator at site $n=0,1,2,\dots,L-1$; $J$ is the nearest-neighbor hopping , $F$ is a constant force, and $l=0,1,2,\dots,N-1$. The on-site potential $\epsilon_n$ is introduced so that the model has a well-defined phase $\phi$, which is needed to calculate the ME; it has no physical significance.

This model has the Stark potential on every $\kappa$-th site and zero otherwise. Since the potential occurs with interval $\kappa$, we can introduce a supercell containing the nearest $\kappa$ sites. If the model contains $N$ supercells, the lattice length will be $L=\kappa N$.

\textit{Mobility edge.}---We will show that this model exhibits exact ME for $\kappa>1$. This can be performed by studying the Lyapunov exponent (LE), which can be obtained exactly by applying Avila's global theory \cite{avila2015global}. We first represent the eigenequation $Eu_n=u_{n+1}+u_{n-1}+\epsilon_n u_n$ (obtained by taking $H|\psi\rangle=E|\psi\rangle$, $|\psi\rangle=\sum_n u_n|n\rangle$) in the form of
\begin{equation}
	\left(\begin{matrix}
		u_{n+1} \\ u_n
	\end{matrix}\right) = T_n
	\left(\begin{matrix}
		u_{n} \\ u_{n-1}
	\end{matrix}\right)
\end{equation}
where the transfer matrix $T_n$ is given by
\begin{eqnarray}\label{transfer}
	T_n &=&
	\left(\begin{matrix}
		E-\epsilon_n & -1 \\
		1 & 0 
	\end{matrix}\right) \nonumber \\
	&=&
	\left(\begin{matrix}
		E-F\kappa l \cos\phi & -1 \\
		1 & 0 
	\end{matrix}\right)
	\left(\begin{matrix}
		E & -1 \\
		1 & 0 
	\end{matrix}\right)^{\kappa-1}.
\end{eqnarray}
Using matrix eigendecomposition, we can express the $(\kappa-1)$-th power of the matrix as
\begin{equation}
	\left(\begin{matrix}
		E & -1 \\
		1 & 0 
	\end{matrix}\right)^{\kappa-1}
	=
	\left(\begin{matrix}
		a_\kappa & -a_{\kappa-1} \\
		a_{\kappa-1} & -a_{\kappa-2}
	\end{matrix}\right),
\end{equation}
with the coefficients
\begin{equation}\label{a}
	a_\kappa = \frac{1}{\sqrt{E^2-4}}\left[\left(\frac{E+\sqrt{E^2-4}}{2}\right)^\kappa - \left(\frac{E-\sqrt{E^2-4}}{2}\right)^\kappa\right].
\end{equation}
 The LE is
\begin{equation}
	\gamma(E) = \lim_{L\to\infty} \frac{1}{2\pi L} \int \ln \|T_L(\phi)\| d\phi \geq 0
\end{equation}
where $T_L = \prod_{n=1}^{L=\kappa N} T_n$ and $\|\cdot\|$ denotes a matrix norm. 

Now, we are going to use Avila's theory to find the LE. Since $T_n(\phi)$ has a holomorphic extension to the neighborhood of $\text{Im}\,\phi$, we can define $T_{n,\varepsilon}(\phi)= T_n(\phi+i\varepsilon)$. By letting $\varepsilon\to\infty$, the transfer matrix becomes
\begin{equation}
	T_l(\phi+i\varepsilon) = \frac{\kappa}{2} l e^{-i\phi}e^{|\varepsilon|}
	\left(\begin{matrix}
		-Fa_\kappa & Fa_{\kappa-1} \\
		0 & 0
	\end{matrix}\right).
\end{equation}
Thus, by a direct computation, we get $\|T_L\|=N!\,|\frac{F}{2} e^{\varepsilon} \kappa a_\kappa|^N$. Within the Stirling's approximation, $\ln N! \approx N \ln N - N$, we have
\begin{equation}
\kappa\gamma_{\varepsilon\to\infty} (E)=\ln\left|\frac{F}{2} a_\kappa\right|+|\varepsilon|+ \ln(\kappa N) - 1.
\end{equation}
According to Avila's global theory, as well as in Refs.~\cite{bourgain2002absolutely,bourgain2002continuity,jitomirskaya1999metal,jitomirskaya2009continuity}, the energy $E$ belongs to a point-like spectrum of $H$, i.e. localized states, iff $\kappa\gamma_\varepsilon(E)>0$. Hence, for $\kappa\gamma_\varepsilon(E)=0$ it is either critical or delocalized (continuous spectrum). The theory also shows that as a function of $\varepsilon$, the Lyapunov exponent is a convex, piecewise linear function, with integer slopes. As we can see in the above, the slope of $\gamma_\varepsilon(E)$ with respect to $\varepsilon$ for $\varepsilon\to\infty$ is exactly $1$. However, in the neighborhood of $\varepsilon\to 0^+$, the slope might be $1$ (when $E$ is in the spectrum) or $0$ (when $E$ is not). Thus, if $E$ lies in the spectrum of the Hamiltonian $H$, we have $\kappa\gamma_\varepsilon(E)=\max\{\ln|\frac{F}{2} a_\kappa| + \ln(\kappa N) - 1,0\}$.

\begin{figure}
	\includegraphics[width=\columnwidth]{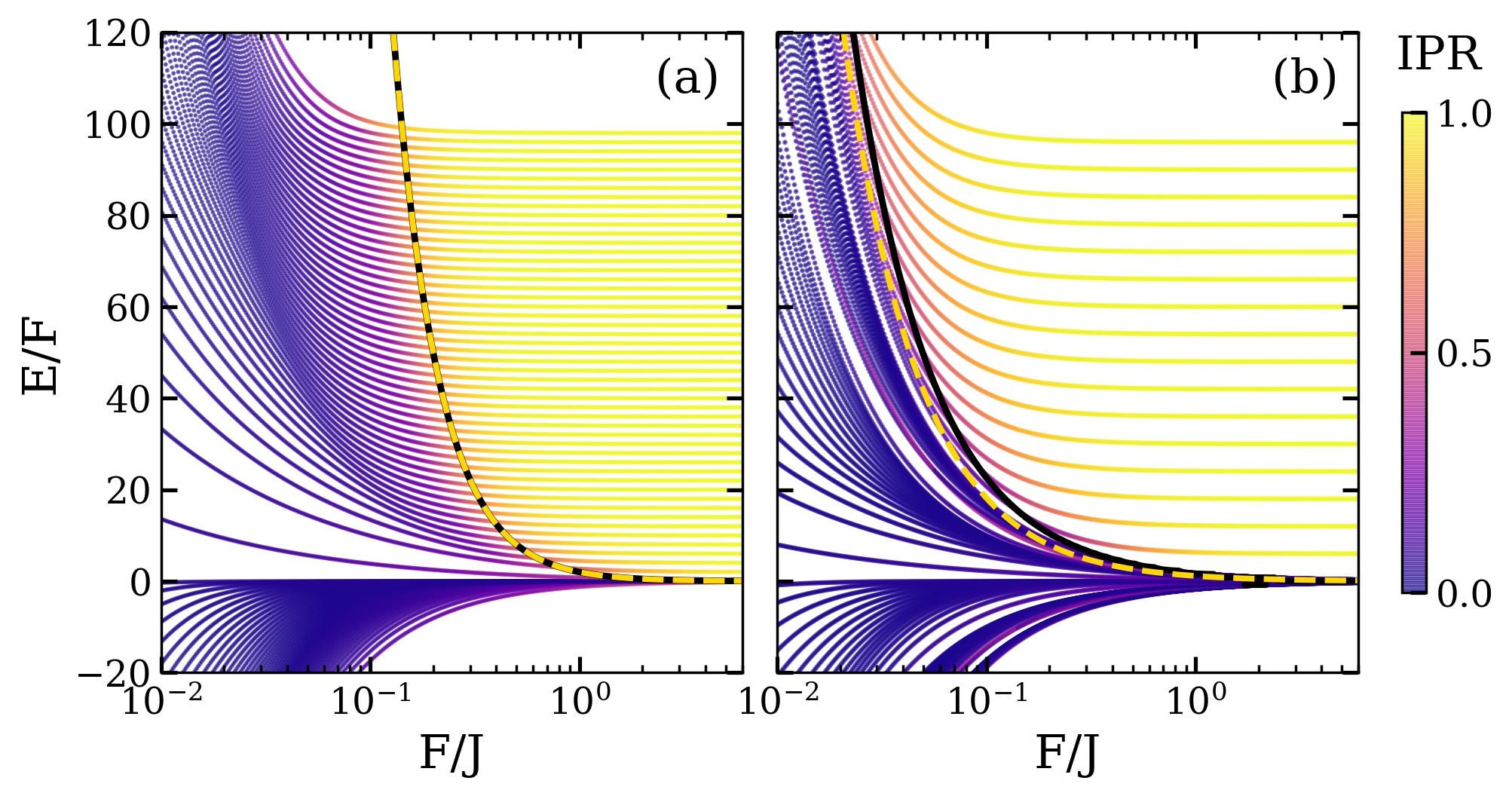}
	\caption{\label{fig:IPR}IPR and energy spectra for $\phi=0$ and $L=100$ (open boundary conditions) with (a) $\kappa=2$ and (b) $\kappa=6$. The black solid curves represent the exact ME from Eq.~(\ref{me}) while yellow dashed curves are the approximate ME, $\tilde E_\text{ME} = (2/F)^{1/(\kappa-1)}$. The exact and approximate MEs coincide for $\kappa=2$.}
\end{figure}

The solution of Eq.~(\ref{transfer}) for $F=0$ behaves like $u_n\sim\exp (\pm \gamma_0(E)n)$ \cite{bentosela1985stark}, which has the localization length of
\begin{equation}
	\xi(E) = \frac{1}{\gamma_0(E)} = \frac{\kappa}{\ln|Fa_\kappa/2|}.
\end{equation}
Note that we have excluded $\ln(\kappa N) - 1$ since it is just a constant, due to the unbounded nature of the potential, and can be absorbed by wavefunction normalization. Reference ~\cite{bentosela1985stark} also shows that the Lyapunov exponent is proportional to $\ln(FN)$. When $|Fa_\kappa|<2$, $\xi\to\infty$ and thus the corresponding eigenstate is delocalized. Hence, the MEs are determined by
\begin{equation}\label{me}
	|Fa_\kappa| = 2.
\end{equation}
This is the central result of this Letter. For instance, for $\kappa=1$ case (the Stark lattice; no ME) the localization transition occurs at $F=2$, whereas for $\kappa=2$ the MEs are $E_\text{ME} = \pm 2/F$ and for $\kappa=3$, $E_\text{ME}~=~\pm~\sqrt{1\pm 2/F}$. In our model, each case only possesses a single ME, which is the ones with the highest $E_\text{ME}$. This is because in this model the energy of localized states is always higher than the extended ones for all $\kappa$ [see Fig. \ref{fig:IPR}].

The ME can also be obtained more simply by using a self-consistent theory \cite{duthie2021self}. That is, by studying the imaginary part of the self-energy obtained from the local Green's function, $G_n(t)=-i\Theta(t)\langle n|e^{-iHt}|n\rangle$. To the first order of the self-energy expansion, $E_\text{ME}$ obeys \cite{duthie2021self}
\begin{equation}\label{duthie}
	\langle\langle \ln[(E_\text{ME}-\epsilon_n)^2]\rangle\rangle - \ln J^2 = 0
\end{equation}
where $\langle\langle \,\cdot\,  \rangle\rangle=\frac{1}{N_0}\sum_{i_0}\frac{1}{2\pi }\int \cdot \, d\phi$ denotes the average over $\phi$ and all the possible end sites $i_0$ (the number of which is $N_0$), e.g. if $\kappa=Q$ then there are $Q$ kinds of end sites: one with the Stark potential, and the $Q-1$ others with zero potential. For $\kappa=2$, we get the same result as the previous method from Avila's theory, 
$E_\text{ME} = 2/F$. Note that for $\kappa>2$, the mosaic model has repeating zero potential of length more than two. This typically demands a higher order theory which can be obtained in a recursive manner from the self-consistent theory
\cite{duthie2021localisation}. However, for our model the first-order approximation, which results in $\tilde E_\text{ME} = (2/F)^{1/(\kappa-1)}$ for any $\kappa>1$, performs relatively well for small $\kappa$'s. This is particularly useful since the exact result [Eq.~(\ref{me})] may require solving a high-order polynomial equation.

\textit{Localization properties.}---We numerically diagonalize the model defined in Eq.~(\ref{model}) for $L=100$ sites with open boundary conditions; $\phi=0$ is set hereafter. The degree of localization of an eigenstate can be quantified using the Inverse Participation Ratio (IPR). The IPR for a $k$-th normalized eigenstate $\psi_k$ is given from the moment
\begin{equation}
	I_q^{(k)} = \sum_n |u_n^{(k)}|^{2q}\propto L^{-D_q(q-1)}
\end{equation}
by choosing $q=2$, whereas $D_q$ is the fractal dimension. For localized eigenstates, $I_q=1$ or $D_q=0$, for an ergodic state, $I_q=0$ (order of $1/L$) or $D_q=1$, and $0<D_q<1$ for fractal states \cite{aoki1986fractal,huckestein1995scaling,janssen1998statistics,de2014anderson}. The IPR and energy spectra for $\kappa=2$ and $\kappa=6$ are shown in Figs. \ref{fig:IPR}(a) and (b), respectively. We superimpose the IPR with the analytical result from Eq.~(\ref{me}) and the approximate one, $\tilde E_\text{ME}$. It can be seen that the exact $E_\text{ME}$ are in excellent agreement with the numerical IPR map, whereas $\tilde E_\text{ME}$ match qualitatively for all $\kappa$ and always underestimate the localization edge for $F/J\ll 1$. For $\kappa=6$, Fig. \ref{fig:IPR}(b), $\tilde E_\text{ME}$ slightly deviates from $E_\text{ME}$, but is still accurate, particularly, for large $F/J$.

\begin{figure}
	\includegraphics[width=\columnwidth]{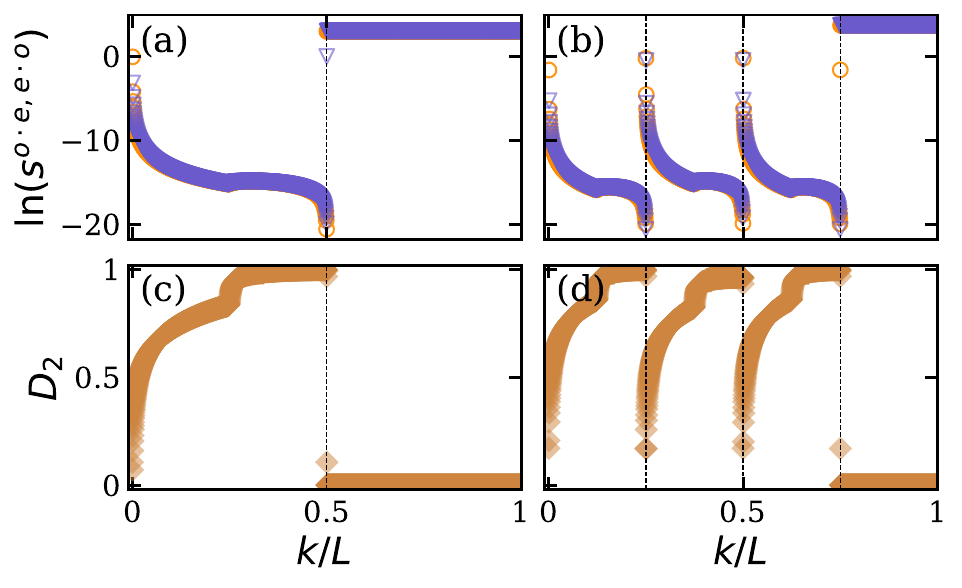}
	\caption{\label{fig:levels}(a) and (b)Level spacing $s^{o-e}$ (blue $\triangledown$) and $s^{e-o}$ (orange $\circ$), and (c) and (d) eigenstate $D_2$ (orange $\Diamond$), for $L=2500$ and a large force ($F=10J$). Columns (a) and ()c): $\kappa=2$ and columns (b) and (d): $\kappa=4$. Vertical dashed lines divide the levels into $\kappa$ equal partitions, in which the rightmost partition contains localized eigenstates. The left $\kappa-1$ partitions, containing $L(1-1/\kappa)$ states, are extended.}
\end{figure}

The important feature of the energy spectra is that localized states appears even with weak forces, $F\ll J$, as opposed to a gigantic electric field corresponding to $F>2J$ in the Stark lattice, and the Wannier-Stark ladder, i.e. equally spaced energy levels and position of the eigenstates, is recovered in high $F/J$ limit, as can be seen clearly in Fig.~\ref{fig:IPR}(b). In this limit, we numerically find that the localized eigenstates form a Wannier-Stark ladder with $E_k^\text{loc}=\kappa F k-(1-1/\kappa)\kappa LF$, whereas the extended states are nearly degenerate with $|E_k|/F\to 0$. The maximum energy of the localized regime grows unbounded in thermodynamic limit since $\max(E_k)=LF$, independent of $\kappa$. There are the highest $\lceil{L/\kappa}\rceil-1$ exponentially localized eigenstates, each occupies a nonzero part of the potential $\epsilon_n$ (one per supercell except for an edge site, $n=0$), with energies matching the potential heights. The lower spectrum contains extended eigenstates, the number of which is $\lfloor(1-1/\kappa)L\rfloor$, and there are $\kappa-1$ gaps [see also the partitions in Figs. \ref{fig:levels}(a) and (b)]. 

A way to discriminate among localized, fractal, and ergodic eigenstates is by analyzing the level spacings of the eigenenergies $E_k$ (arranged in increasing order), which are the odd-even (even-odd) spacings $s_k^{o-e}=E_{2k+1}-E_{2k}$ ($s_k^{e-o}=E_{2k} - E_{2k-1}$). Localized levels will have overlapping $s_k^{o-e}$ and $s_k^{e-o}$ values, whereas the ergodic levels are doubly degenerate with a gap \cite{aubry1980analyticity}; fractal states have strongly scattered levels. Figures \ref{fig:levels}(a) and (b), and the corresponding $D_2$ in Fig. \ref{fig:levels}(c) and (d), highlight the appearance of $\kappa$ equal partitions of the eigenstates in the level spacings for every case. The localized states are located at the rightmost partition. All the extended eigenstates are degenerate except at partition edges. Every partitions edges corresponds to the spectral gaps in Fig. \ref{fig:IPR}. In addition, numerical study indicates that in thermodynamic limit, $L\to\infty$, these gaps are persistent. 

The extended states, with energies $E_{k\leq\lfloor(1-1/\kappa)L\rfloor}$, are nonergodic since $0<D_2<1$; particularly the ones at partition edges. The level spacing is neither gapped nor scattered, which indicates the absence of fractality that typically occurs in disordered or an incommensurate lattice. This pecularity is related to the continuity of Bessel function, $\mathcal J_{n-m}(F/2J)$, as the solution for $\kappa=1$. There are no sharp transition as $F\to 2J$ and become localized to one site. For our mosaic lattice, one can regard the low energy wavefunction as a combination of Bessel-like amplitude and solution of a regular lattice with a Dirac comb potential, although the exact solution is not found at present. We numerically find $\kappa-1$ blocklike transitions of $D_2$ in Figs. \ref{fig:levels}(c) and \ref{fig:levels}(d). Every eigenstate in the partition edges is localized within the leftmost supercell (not shown). States with higher energies are increasingly spread out until they fills the entire lattice ($D_2=1$) for the state just before the next partition edge. Note that another model with nonergodic extended states do exist, such as the generalized Rosenzweig-Porter (GRP) model \cite{kravtsov2015random}. In contrast to the nearest-neighbor hopping used in our model, GRP contains random long-range hopping.

\begin{figure}
	\includegraphics[width=\columnwidth]{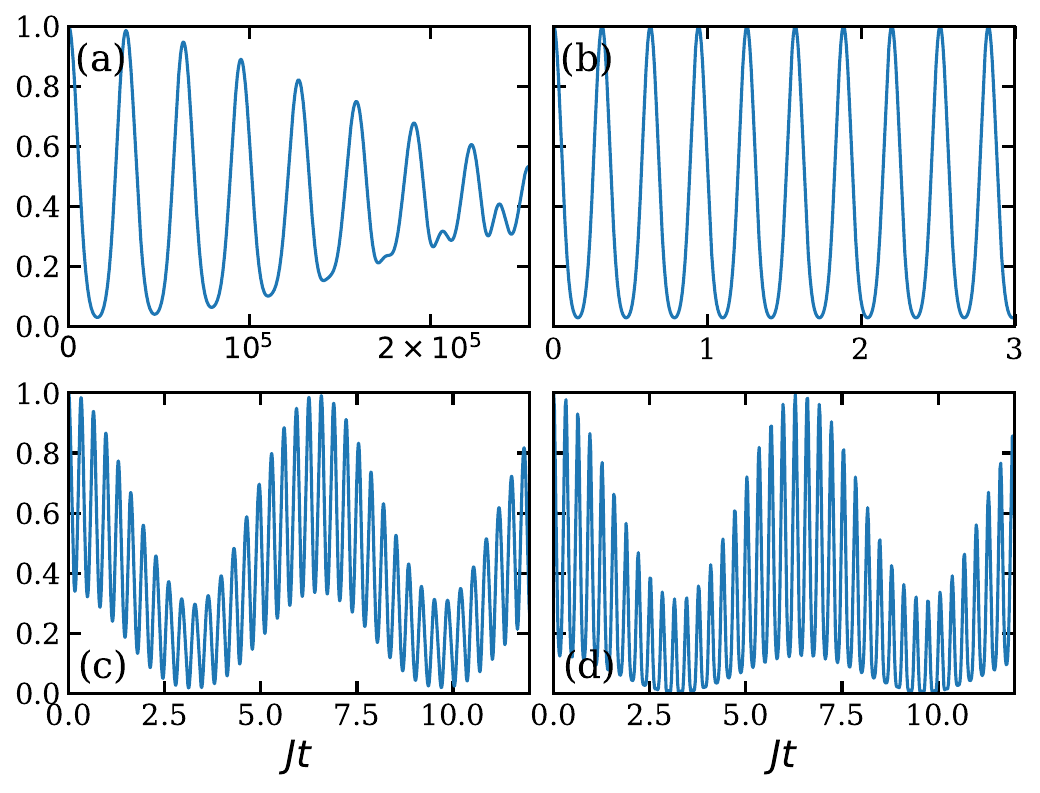}
	\caption{\label{fig:f_t}Survival probability $f(t)$ ($L=100, F/J=10$, and $\kappa=2$) for the wave packet centered at (a) $k_0 = k_\text{loc} - L/4$ (extended), (b) $k_0 = k_\text{loc}+L/4$ (localized), (c) $k_0 = k_\text{loc}$ (crosses ME), and (d) analytical approximation for (c). Here $k_\text{loc} = 50$. }
\end{figure}

\textit{Excitation dynamics.}---To dynamically probe the existence of ME, we study the time dependence of a Gaussian wave packet centered at $E_{k_0}$, that is, $|\Psi(0)\rangle \propto \sum_k \exp\left[-i(k-k_0)^2/2\sigma^2\right]|\psi_k\rangle$ up to a normalization constant. We calculate the survival probability \cite{torres2018generic},
\begin{equation}
	f(t)=\left|\langle\Psi(t)|\Psi(0)\rangle \right|^2 = \left| \sum_k \left| \langle\psi_k|\Psi(0)\rangle\right|^2 e^{-iE_kt}\right|^2,
\end{equation}
which is the probability of finding the initial state after time $t$. We can analytically calculate $f(t)$ by considering the thermodynamic limit and dividing the summation into extended ($k \leq k_\text{loc}$) and localized ($k > k_\text{loc}$) states where $k_\text{loc}=\left\lfloor(1-1/\kappa\right)L\rfloor$. For localized wave packets $\Psi(0)$, we assume that $k_0$ is deep in the Wannier-Stark ladder, with $E_k^\text{loc}$, so that the summation extends to infinity, and small $\sigma^2$. To see the effect of ME, we allow the tail to cross ME, $\sum_{k\leq k_\text{loc}}|\langle\psi_k|\Psi(0)\rangle|^2 \approx \Delta$, and with a common energy $\varepsilon\ll F$. This approximation is possible because the extended spectrum is effectively degenerate (average energy $|E|/F\sim10^{-3}$ for $F=10J$ and $\kappa=2$) except for the states near the partition edges ($|E|/F\sim 0.1$ at maximum). Thus, we get $f(t)$ proportional to
\begin{equation}\label{exact}
\Delta^2 + \vartheta_3 \left(\frac{\kappa Ft}{2},e^{-\alpha}\right)^2 + 2\Delta \vartheta_3\left(\frac{\kappa Ft}{2},e^{-\alpha}\right)\cos\,(\theta+\varepsilon)t,
\end{equation}
where $\alpha=1/\sigma^2$, $\theta(k_0)=\left(k_0 - k_\text{loc}\right)\kappa F$, and 
\[\vartheta_3(z,q) = \vartheta_3(z+\pi,q) = \sum_{s=-\infty}^{\infty} q^{s^2} e^{2isz}\]
is the Jacobi theta function \cite{whittaker2021course} with $s=k-k_0$. For $\Delta=0$ (no ME crossing; localized), $f(t)$ becomes periodic with $T=2\pi/\kappa F$, independent of $k_0$, and Eq.~(\ref{exact}) becomes exact.

Figures \ref{fig:f_t}(a)-\ref{fig:f_t}(c) show $f(t)$ for the wave packet in extended, localized, and ME crossing regimes, respectively. Clearly the extended and localized cases oscillate with separate timescales. For the wave packet crossing the ME [Fig. \ref{fig:f_t}(c)], we choose $k_0=k_\text{loc}$, resulting in an oscillation with the same frequency as in localized case, but bounded between multiple envelopes. This can be approached analytically by taking $\Delta=0.5$ and $\varepsilon=E_{k_\text{loc}}\approx 0.1F$ since it is significantly larger than the rest of nearby extended spectrum. The result is in Fig. \ref{fig:f_t}(d), which shares qualitative features with Fig. \ref{fig:f_t}(c).

\textit{Conclusions.}---In this Letter, we have analytically demonstrated the existence of ME in a disorder-free lattice. We show that the analytical ME is in excellent agreement with the numerical localization properties. For $\kappa>1$, localized states exist even with weak fields. The existence of the mobility edge may also probed by the time evolution of an initial wave packet. The survival probability oscillates with separate timescales for extended and localized regimes. Our Letter suggests the possible existence of the disorder-free mobility edge in the many-body localized Stark lattice \cite{schulz2019stark} with mosaic spacings.

F.P.Z. thanks the Ministry of Higher Education and Research of Indonesia for Research Funding 2021. The numerical results were obtained using code written in \verb+NumPy+ \cite{harris2020array} and \verb+QuTiP+ \cite{johansson2013qutip}, and the figures were made using \verb+matplotlib+ \cite{hunter2007matplotlib}. 


\bibliography{dwiputra} 

\clearpage

\end{document}